\documentclass[]{emulateapj}
\usepackage{amsmath}
\usepackage{amssymb}
\usepackage{amstext}
\usepackage{graphicx}
\usepackage{epsfig}

\usepackage{color}
\definecolor{myColor}{rgb}{0.9,0.9,0.9}

\begin{document}

\submitted{Submitted to ApJ on September 28, 2007}
\title{Production of Millisecond Dips in Sco X-1 Count Rates by
 Dead Time Effects}

\author{T. A. Jones\altaffilmark{1}, A. M. Levine \altaffilmark{2},
 E. H. Morgan \altaffilmark{2}, and S. Rappaport\altaffilmark{1}}   

\altaffiltext{1}{Department of Physics and Kavli Institute for Astrophysics
 and Space Research, MIT, Cambridge, MA 02139; {\tt }}

\altaffiltext{2}{Kavli Institute for Astrophysics and Space Research,
 MIT, Cambridge, MA 02139; {\tt }}


\begin{abstract}

\citet{chang06} reported millisecond duration dips in the X-ray
intensity of Sco X-1 and attributed them to occultations of the source
by small trans-Neptunian objects (TNOs). We have found multiple lines
of evidence that these dips are not astronomical in origin, but rather
the result of high-energy charged particle events in the {\it RXTE}
PCA detectors. Our analysis of the {\it RXTE} data indicates that at
most 10\% of the observed dips in Sco X-1 could be due to occultations
by TNOs, and, furthermore, we find no positive or supporting evidence
for any of them being due to TNOs.  We therefore believe that it is a
mistake to conclude that any TNOs have been detected via occultation
of Sco X-1.

\end{abstract}


\section{Introduction}
\label{sec:intro}

\citet{chang06} found statistically significant 1-2 millisecond
duration dips in the count rate during X-ray observations of the
bright X-ray source Sco X-1 carried out with the Proportional Counter
Array (PCA) on the {\it Rossi X-ray Timing Explorer} ({\it RXTE}) and
attributed them to occultations of the source by small objects
orbiting the Sun beyond the orbit of Neptune, i.e., trans-Neptunian
objects (TNOs).  In all, \citet{chang06} found some 58 dips in
approximately 322 ks of Sco X-1 observations.  Given that the {\it
RXTE} spacecraft moves through the diffraction-widened shadows of any
TNOs at a velocity of $\sim30$ km s$^{-1}$, dips of $\sim$2 ms
duration should correspond to a TNO size of $\sim$60 m.  If the
identification of these dips with occultations by TNOs is correct, the
dips would provide extremely valuable information on the number and
distribution of solar system objects of $\sim20$-100 m in size.  We
have found evidence that these dips are produced by electronic
dead-time as a result of high-energy charged particle events in the
RXTE PCA detectors.  Preliminary reports of our results were given by
\citet{tajatel, tajast}; herein we give a more detailed and complete
report.

\section{Average Properties of Dips in the Sco X-1 Count Rates}
\label{sec:data}

\begin{figure}[htb]
\plotone{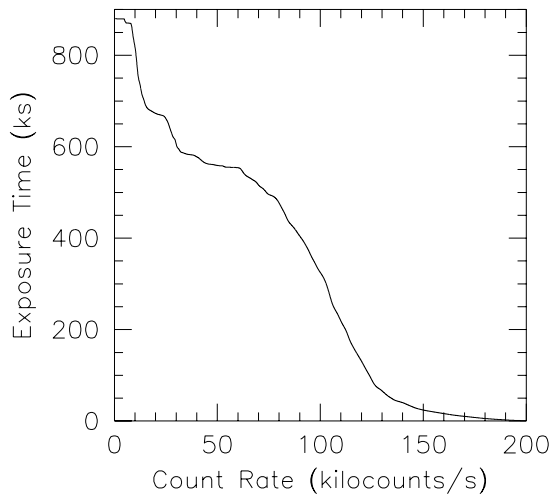}
\caption{Cumulative distribution of exposure time as a function of PCA
count rate during observations of Sco X-1 which yielded high time
resolution data and were used for the present analysis.  The ordinate
gives the total exposure time at count rates greater than that of the
corresponding rate on the abscissa.  Two-LLD events have been included
in the count rates when available (see text).}
\label{fig_pcaexpos}
\end{figure}

Subsequent to the report by \citet{chang06}, we searched for dips of
the type they describe in the archival data obtained in $\sim 880$ ks
of {\it RXTE}/PCA observations of Sco X-1. These observations were
performed starting in early 1996 soon after the launch of the
spacecraft.  For the search we only used observations that provided
count rates of events with a time resolution of $\lesssim 0.25$ ms.
The total count rate depends on the strength of Sco X-1 at the time of
observation as well as on the number of Proportional Counter Units
(PCUs) that were operating and the location of the source within the
field of view; it ranges from below 40,000 cts s$^{-1}$ to nearly
200,000 cts s$^{-1}$ (see Figure~\ref{fig_pcaexpos}).  

Coincidences within a $\sim10$~$\mu$s window among two or more of the
measurement chains in a Proportional Counter Unit (PCU) are normally
used to identify charged particle events \citep[see][and references
therein, for technical information on the PCA]{pca06}.  However, the
intensity of Sco X-1 is so high that there is a substantial count rate
due to the detection of two X-ray photons in two distinct regions of
the detector serviced by different measurement chains within the
10~$\mu$s window.  For most of the Sco X-1 observations, the rates of
such so-called two lower-level discriminator (``2-LLD'') events were
telemetered with sub-millisecond time resolution.  In those cases
where the 2-LLD event data are available, we add 2 counts for each
2-LLD event to the counts of single LLD events.

In our search, we looked for instances when the number of counts in a
time ``window'' that is an integral multiple of 0.25 ms ($2^{-12}$~s
to be precise) was less than that reasonably expected to occur given
the mean count rate.  The search was performed for each of the seven
window intervals of 0.75, 1, 1.5, 2, 2.5, 3, and 4 ms. The expected
number of counts in each window was determined from the running
average count rate in a time interval centered on the time window of
128 ms for the 0.75 and 1 ms windows, 192 ms for the 1.5 ms window, or
256 ms for the 2, 2.5, 3, and 4 ms windows.

Given $N$, the number of counts actually measured in the window, and
$N_{exp}$, the expected number of counts in that window based on the
running average, we computed $P(n \leq N|N_{exp})$, the probability
that $N$ or fewer counts would be detected based on the simplistic
assumption that the counts obey a Gaussian distribution with mean
$N_{exp}$ and standard deviation $\sqrt{N_{exp}}$.  We define the
detection of a dip as any instance in which we found $P(n \leq
N|N_{exp}) < 10^{-10}$.  In the search, the same dip could be found
multiple times, e.g., in contiguous intervals when it was longer than
the window interval or for different window intervals.  We generated a
list of 203 dips in which these duplications had been eliminated.  Of
these, 196 occur in data in which 2-LLD events had been taken into
account. All but three of the 58 dips of \citet{chang06} and all but
14 of the 107 dips identified by \citet{chang07} were identified in
our search. Eight of the latter 14 dips were not found by us because
we did not search 5 orbits of data that \citet{chang07} searched, and
because 6 of the 14 did not meet our significance threshold (perhaps
because we included 2-LLD events).  We detected more dips than
\citet{chang07} because we included 2-LLD events, because our
significance threshold was slightly lower, and possibly because of
other minor differences in the searches.
 
The frequency of occurrence of dips was found to be a function of
count rate, and, as one should expect, tended to be higher at the
higher count rates.  Given our detection criteria, we found no dips at
count rates below 43,000 cts s$^{-1}$ and only three at count rates
below 55,000 cts s$^{-1}$.  We searched for dips in data which
included approximately 570 ks of observations in which the count rates
were above 55,000 cts s$^{-1}$ (Fig.~\ref{fig_pcaexpos}).  For each of
the seven window durations, we computed $N$ and $N_{exp}$ for window
intervals that started every 0.25 ms. Thus the number of independent
trials must be less than $7 \times 570,000/0.00025 \sim 1.6 \times
10^{10}$.  If the actual probability of finding a dip due to a
statistical fluctuation in each trial is $\lesssim 10^{-10}$, then at
most a few of the 203 dips could be the result of statistical
fluctuations.

In every case, the actual probability of getting $N$ or fewer counts
differs from our computed value for at least two reasons.  First, the
Gaussian distribution we utilized overestimates the probabilities of
small numbers of counts by large factors when compared with a Poisson
distribution of the same mean.  Second, the intensity of Sco X-1 is
time variable, and there is some chance that the mean intensity at the
time of a dip was lower than that corresponding to $N_{exp}$.  If
source variability was important, our detection procedure would have
led to numerous spurious detections, particularly for the longer
window durations.  However, only four dip-like events were most
significantly detected when using either the 3 or 4 ms window
durations.  We further checked the effects of source variability by,
first, integrating properly normalized power density spectra of Sco
X-1 count rate data over the frequency range of 5-300 Hz, 4-250 Hz,
4-200 Hz, 4-170 Hz, or 4-125 Hz for the 1.5, 2, 2.5, 3, and 4 ms
window durations, respectively. The resulting root-mean-square
fractional amplitudes in the given frequency bands are typically
$\sim2.5$\% and are almost always less than $\sim4.5$\%.  These
variablity estimates for each of the five window durations were then
used together with the average count rates on an ({\it RXTE})
orbit-by-orbit basis to estimate the effects on the dip detection
probabilities.  The results are consistent with the paucity of dips
detected most significantly using the 3 ms and 4 ms window durations
and indicate that statistical fluctuations in the presence of source
variability did not produce more than $\sim5$ spurious detections.

\begin{figure}[t]
\plotone{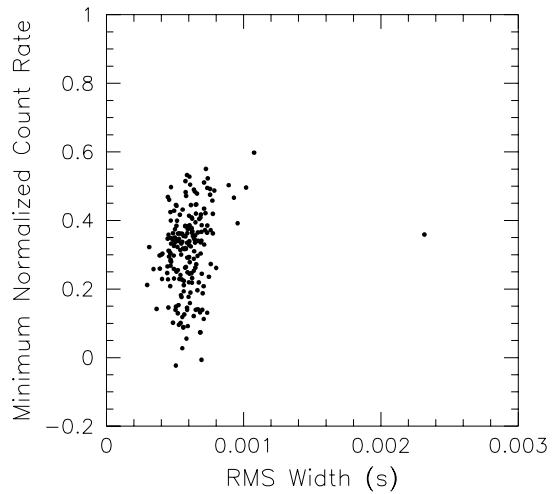}
\caption[Width and depths for detected dips]{Distribution of fitted
 RMS widths and minimum normalized count rates for 203 detected dip
 events.  The minimum normalized count rate is the value of the model
 count rate at the center of the fitted Gaussian divided by the
 average count rate away from the dip.}
\label{fig_realwd}
\end{figure}

In order to obtain estimates of dip widths and depths, we used the IDL
procedure {\it gaussfit} to fit each dip profile with a function
$f(t)$ representing a constant count rate plus a Gaussian shaped dip:
\begin{equation}
f(t) = A - B e^{-(t - t_0)^2/2 \sigma^2}
\end{equation}
where $A$, $B$, $t_0$, and $\sigma$ were the parameters to be
determined.  The values of the fitted widths as parameterized by the
values of $\sigma$, and of the fitted minimum normalized count rates,
i.e., the values of $1 - B/A$, of the $\sim 200$ dips are shown in
Figure~\ref{fig_realwd}.  The key features of this figure are the
relatively narrow range of widths and absence of long dips. Nearly all
of the events have $0.4 \leq \sigma \leq 0.8$ ms, and there is only
one dip with $\sigma > 1.1$ ms.  These aspects of the plot are
discussed below (see Section~\ref{sec:disc}).

If the counting rate dips are the result of occultations, then we
would expect that diffraction effects would produce small count rate
increases on either side of the dips. The sizes of these sidelobes and
indeed the other details of the dip profiles depend on a number of
factors including the sizes, shapes, distances, and velocities of the
occulting bodies, the impact parameters characterizing the occultation
events, and the velocities of the {\it RXTE} spacecraft at the times
the dips were observed.

To quantify our expectations of the average diffraction sidelobe size
and other typical properties of dips produced by occultations by TNOs,
we performed a Monte Carlo computer simulation of an ensemble of
occultation events in which the occulting bodies were assumed to be
opaque spheres, to have radii $s$ that follow a distribution
$\frac{dN}{ds} \sim s^{-4}$, and to follow prograde circular orbits at
a distance of 40 AU from the Sun.  Since the effects of diffraction
are wavelength dependent, we took the spectrum of detected X-rays to
be that of a typical pulse height spectrum of Sco X-1 measured with
the PCA.  The simulation was carried out for each of four cases in
which the relative velocity between the spacecraft and the shadows of
the occulting bodies was a fixed value, viz. 15, 25, 30, or 35 km
$s^{-1}$.

Simulated dip profiles were computed for a large number of
occultations with object size and occultation impact parameter chosen
at random according to the appropriate distributions. The profiles
were normalized to unity at times far from the occultation centers and
were then inserted into the real PCA count rate data by multiplying
the actual count rates by the dip profiles.  The dip center times were
chosen at random among the 880 ks of observations that were used for
the present analysis.  These data were then searched for dips using
the search algorithm described above.

\begin{figure}[t]
\plotone{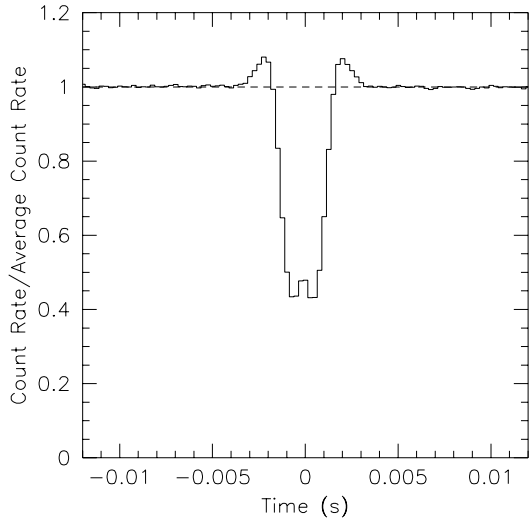}
\caption[Average simulated inserted dip lightcurves]{Average profile
of the 2189 detected 25 km/s model population occultation events.  The
profiles are superposed after centering and stretching according to
the parameters of the fitted Gaussian curves and have been normalized
to the count rates at more than 0.007 s from the dip center.}
\label{simprof}
\end{figure}

We fit the detected model dips with the function given by eq. (1). The
light curves of the PCA data containing the detected simulated dips
were superposed after aligning the dip centroids and rescaling the
time scale of each light curve in order to normalize all of the fitted
widths to the value $\sigma = 0.85$ ms.  Interpolation of the rescaled
bin times to 0.25 ms time bins was required to superpose the rescaled
light curves, and, as a consequence, adjacent bins in the
superposition are not completely statistically independent.  The
results are shown in Figure~\ref{simprof}.

\begin{figure}[t]
\plotone{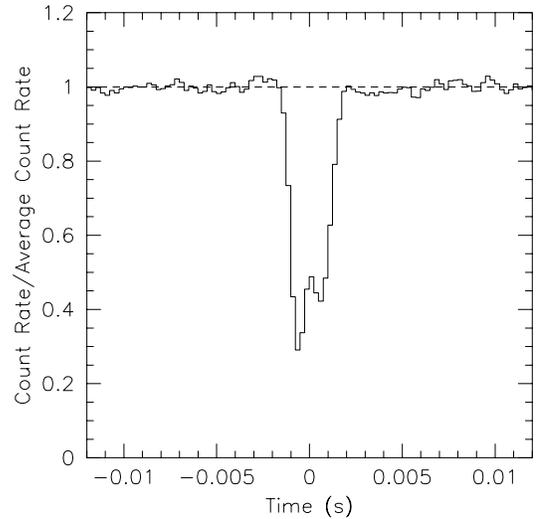}
\caption[Average lightcurve]{Average dip profile of 202 dip events,
 normalized to the mean count rate at more than 0.007 s from the dip
 center (dashed line). The standard deviation computed for these bins
 is 0.016 (1.6\% of the count rate).}
\label{fig_superpos_202}
\end{figure}

\begin{figure}[b]
\plotone{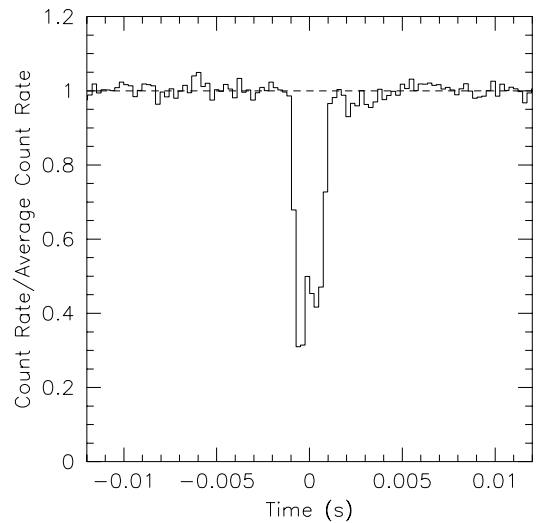}
\caption[Superposition of 109 events of similar width]{Superposition
 of all 109 events with width parameters between 0.55 and 0.75 ms,
 with 0.25 ms time bins, normalized to the mean count rate at more
 than 0.007 s from the dip center (dashed line).}
\label{fig_superpos_109}
\end{figure}

The average profile of the real detected dips was similarly
constructed by superposing the PCA light curves of 202 of the 203
detected dips after alignment and stretching; the one dip with width
$\sigma \sim 2.3$ ms was excluded. The results are shown in
Figure~\ref{fig_superpos_202}.  To check this, we also superposed only
those 109 dips with fitted widths in a narrow range, i.e., with $0.55$
ms $ < \sigma < 0.75$ ms.  The light curves containing those 109 were
aligned and summed, but no stretching was done.  The results are shown in
Figure~\ref{fig_superpos_109}.

Per the results shown in Fig.~\ref{simprof}, we expected to see
sidelobes with intensities as high as $\sim$8\% above the mean count
rate determined substantially away from the superposed dips.  The
superposed light curves do not exhibit diffraction sidelobes as high
as those evident in the average profile of the simulated dips, despite
having statistics sufficient to reduce fluctuations to $\sim$1.6\%
(1~$\sigma$) of the mean count rate.  While the differences are not
sufficiently significant to be conclusive, they strongly suggest there
may be a problem with the occultation hypothesis. We elaborate on this
in the discussion section below.

Three additional potential problems with the occultation
interpretation are manifest from the dip profiles.  First, the summed
dip profile is distinctly asymmetric in shape as \citet{chang06}
suggested for many of the individual dips.  Second, the distribution
of dip widths is narrower than what one would expect from occultations
by bodies with a power-law size distribution of index -4, i.e., there
are fewer than expected statistically significant dips with Gaussian
FWHM widths greater than $\sim1$ ms.  Third, when diffraction effects
are taken into consideration, one would expect to see a correlation
between the fitted widths and the fitted minimum count rates, such
that longer dips tend to be deeper on average.  No such correlation is
seen in Figure~\ref{fig_realwd}.

\begin{figure*}   
\epsscale{0.85}
\plotone{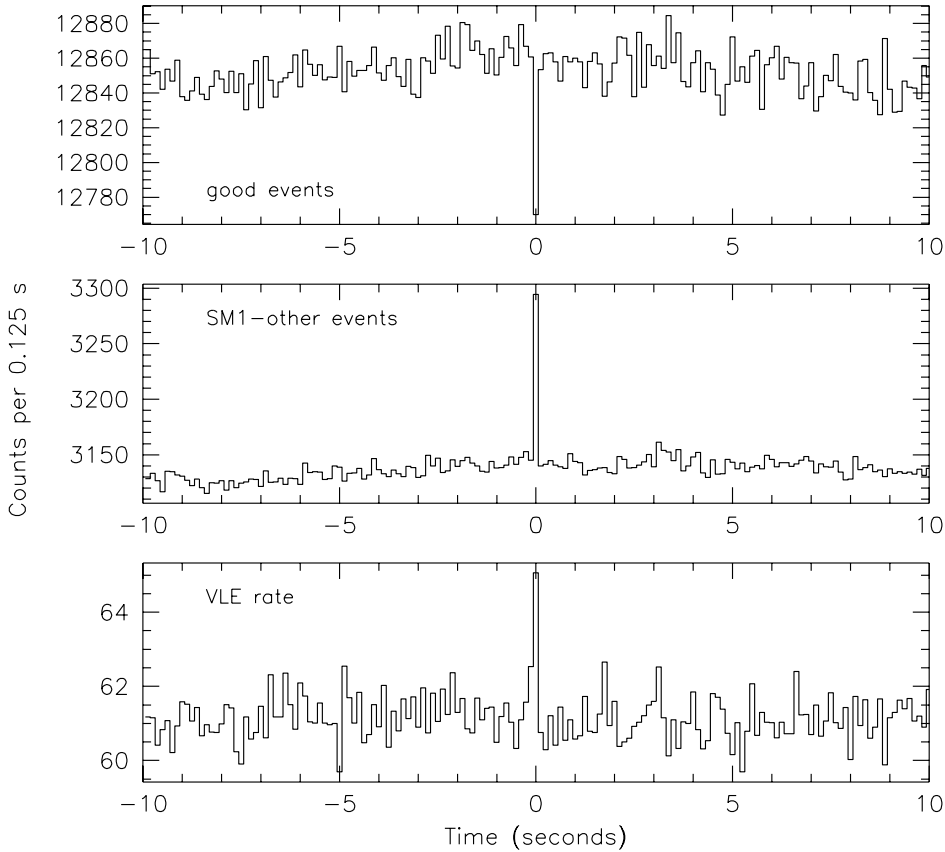}
\caption{Counts per 1/8-s time bin of different types of PCA detector
events superposed, i.e., averaged, around the times of 201 dips.  The
data used for this figure were recorded in Standard Mode 1.  The
superposition was accomplished such that the bin at $time = 0$~s
includes the identified dips.  The counts include events from all
(typically 3 to 5) of the operating PCUs. Top panel: good xenon
counting rate data. The decrease in counting rate due to the dips is
apparent.  The small ($\sim$0.7\%) drop in the counting rate is
explained in the text.  Middle panel: counting rate data of SM1-other
events that include multiple-LLD events (see text). A highly
significant enhancement in the vicinity of the dips is evident.
Bottom panel: VLE event rate data superposed around the dip times,
also showing a statistically significant peak.  Note that the peak in
the VLE event rate is approximately one VLE event per detector per dip
event (i.e., $\sim$4 excess events per dip).}

\label{fig:xraylc}
\end{figure*}
  
\section{Search for an Alternate Explanation}
\label{sec:expl}

These findings prompted us to further explore alternative explanations
for the dips. Only one hypothesis appeared to be worthwhile to pursue,
i.e., that the dips are caused by electronic dead time in response to
some type of charged particle shower in the spacecraft.
Unfortunately, no information with millisecond time resolution was
available on the non-X-ray background during the Sco X-1 observations.
Counts of good events, very large events (VLEs), propane-layer events,
and a catch-all category of other types of events (hereafter SM1-other
events) that includes multiple LLD events are available at 1/8 s time
resolution from Standard Mode 1; most other types of data are only
available with 16-s time resolution.

The VLE flag for a PCU is set when the electronics detect an event in
that PCU with energy greater than $\sim$100 keV; this can happen in
response to the ionization produced by a single charged particle or to
that produced by multiple charged particles which penetrate the
detector nearly simultaneously. Such a large event can produce ringing
in the front-end of an electronic measurement chain.  Therefore, in
response to the occurrence of a VLE, the digital logic shuts down the
electronic processing of events in that PCU for a fixed time period,
chosen to be 50 $\mu$s for almost all of these Sco X-1
observations. In addition, each of the 6 main xenon-layer measuring
chains is disabled until its charge drops to an acceptable level. In
order for the detector to be shut down for an extended period ($>$
1~ms), an extraordinary amount of charge must be deposited on most of
the 6 main measuring chains; it is unclear, at present, whether this
can happen in response to a single charged particle.

Figure~\ref{fig:xraylc} shows counts of three different types of
events from Standard Mode 1 in 1/8-s time bins superposed around the
times of 201 dips.  In each panel, the centers of the dips have been
placed in the bin at $time = 0$.  The top panel shows the rates of
good events, i.e., those not identified as being due to charged
particles, in the main xenon layers of all operating PCUs, and clearly
shows the superposed dips; two-LLD events are not included in these
rates.  The counting rate drops by only $\sim$0.7\% because of the
dilution of a $\sim2$ ms dip within a 128 ms bin.  In contrast, the
middle panel shows the {\em enhancement} of the counting rate of
SM1-other events in the vicinity of the dips.  The peak is highly
significant ($\sim$38 $\sigma$).  The bottom panel corresponds to the
VLE event rate superposed around the dip times.  This peak is also
statistically very significant ($\sim$7 $\sigma$).  The increase in
the VLE rate is nearly so large as to be consistent with the detection
of $\sim$1 VLE per PCU per dip.  We discuss this further below.

The enhancements in the SM1-other event and VLE rates around the times
of the dips indicate that there is an increase in the rate of
detection of non-X-ray events.  We speculate that these non-X-ray
events interrupt normal event processing for 1-2 milliseconds in most
of the PCUs roughly once per hour due to the collection of very large
amounts of charge. Such an energetic event may be the consequence of a
particle shower produced by the collision of a high-energy cosmic ray
with a nucleus in the {\it RXTE} spacecraft. In any case, further
clarification of the causes of the observed dips would be of interest.

\section{Discussion}
\label{sec:disc}

The observed dips have widths and depths that are approximately what
one might expect to be produced by occultations by TNOs, even though
much wider dips would be detectable in principle (given appropriate
depths).  Thus we are obligated to seriously consider the hypothesis
that some or all of the observed dips are the product of TNO
occultations.  However, close examination of the {\em RXTE} PCA data
reveals six signatures that independently indicate that few and
possibly none of the observed dips are due to occultations by TNOs.
The signatures are (1) the numbers of SM1-other events during the
dips; (2) the numbers of VLE events during the dips; (3) the absence
of the expected diffraction sidelobes; (4) the temporal asymmetry of
the dips; (5) the almost total lack of dips longer than $\sim$1 ms;
and (6) the lack of correlation between dip duration and depth.  We
discuss each of these in turn.

\begin{figure*}   
\epsscale{0.85}
\plotone{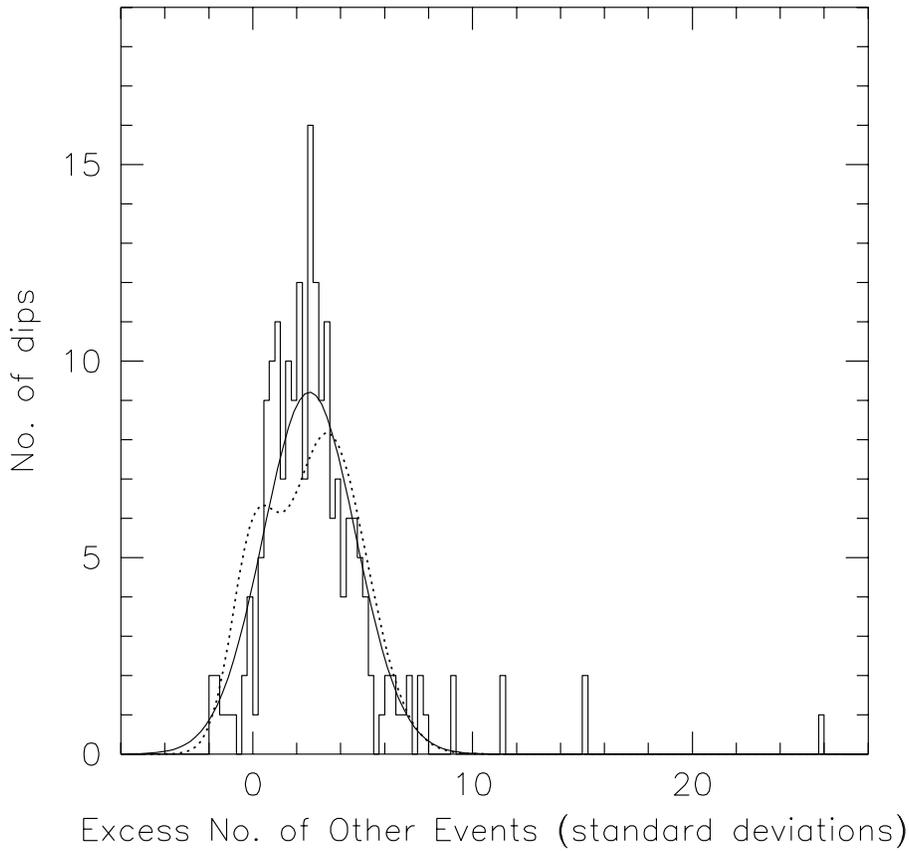}
\caption{Histogram of the numbers of excess SM1-other events in 1/8-s
time bins corresponding to the times of 201 of the 203 dips.  The
number of excess SM1-other events is given in terms of the square root of
the mean number of SM1-other events per bin away from the time of the
dip.  The solid smooth curve represents the best fit with $f_{\rm occ} =
0$, $\sigma_{cr} = 2.1$ and the dashed curve represents a formally
{\em un}acceptable fit with $f_{\rm occ} = 0.24$, $\sigma_{cr} = 1.8$ (see
text). }
\label{fig:otherhist}
\end{figure*}

(1) {\bf SM1-Other Events:} Fig.~\ref{fig:xraylc} shows that there is,
on average for 201 dips, a large excess of SM1-other events at or near
the times of the dips. On average, individual dips should show an
excess of SM1-other events at the $2.7\, \sigma$ level.  In
Fig.~\ref{fig:otherhist} we show a histogram of the number of
SM1-other events in the 1/8-s time bins corresponding to the dips
expressed in standard deviations from the mean.  The mean value
obtained from the histogram is $\sim2.7\, \sigma$, as expected.

If one makes the reasonable assumption that the numbers of SM1-other
events should not be affected by true occultations (other than by
negligible increases due to reductions in the electronic dead time),
then one may estimate the maximum fraction, $f_{\rm occ}$, of the observed
dips that represent genuine occultation events that is consistent with
this distribution of SM1-other events.  We constructed the following
simple function with which to fit the histogram, and thereby constrain
$f_{\rm occ}$:
\begin{equation}
{\rm probability} = \frac{1}{\sqrt{2\pi}}f_{\rm occ} 
e^{-C^2/2} + \frac{1}{\sqrt{2\pi \sigma_{cr}^2}}(1-f_{\rm occ})e^{-(C- 
\overline{C}_{cr})^2/2\sigma_{cr}^2}
\end{equation}
where $C$ is the number of excess SM1-other counts (in units of
standard deviations of the counts per bin in each PCA light curve),
and $\overline{C}_{cr}$ is the mean of $C$ for those dips which are
not the results of occultation events and which we take to be
$\approx2.7/(1-f_{\rm occ})$.  The distribution of the numbers of excess
SM1-other counts is wider than what would be expected from a Poisson
distribution with a mean equal to the slightly increased (on average)
number of SM1-other events per bin; the width of this component of the
fitting function is adjusted by means of the parameter $\sigma_{cr}$.
Fits of the function to the histogram in Fig.~\ref{fig:otherhist} were
carried out with $\chi^2$ fits using both Gaussian and Cash (1979)
statistics.  If we neglect the tail of the distribution at high
numbers of excess SM1-other events, i.e., at $>9\sigma$, we obtain
formally acceptable fits with values of $f_{\rm occ}$ in the range 0.0 to
0.12 and values of $\sigma_{cr}$ in the range 1.85 to 2.35 (based on
Gaussian statistics; the limits represent the formal joint 95\%
confidence range). Using Cash statistics, we obtain formally
acceptable fits with values of $f_{\rm occ}$ in the range 0.0 to 0.11 and
values of $\sigma_{cr}$ in the range 1.65 to 2.15 (95\% confidence).
These results indicate that fewer than 11\% of the 203 dips might be
the product of TNO occultations.

(2) {\bf VLE events:} Figure~\ref{fig:xraylc} also shows that there is
an excess of VLE events around the times of the dips.  The difference
between the background rate and that in the 1/8-s bin containing the
dips is very close to 4 (actually $3.9 \pm 0.5$) extra VLE events per
dip.  The peak in Fig.~\ref{fig:xraylc} is significant at the $7
\sigma$ level.  If there is precisely one VLE per operating PCU for
each non-TNO dip, then we would expect on average an excess of 4.67
VLEs per non-TNO dip. If only the non-TNO dips contribute to the
excess VLE events then there is an upper limit to $f_{\rm occ}$ that is
consistent with the observations.  If we further allow that the
statistical mean excess number of VLE events per dip may have been as
small as 2.9, then a simple calculation gives the limit $f_{\rm occ}
\lesssim 0.38$ (95\% confidence). This limit is weaker than for the
SM1-other events, and, furthermore, is compromised by the possibility
that more than one VLE event could be produced in a operating PCU in a
single cosmic-ray induced dip.

(3) {\bf Lack of diffraction sidelobes:} In Fig.~\ref{simprof} we
showed an average profile of simulated dips that had been inserted
into actual PCA data.  It should be compared to averages of the actual
measured dip profiles in Figs.~\ref{fig_superpos_202} and
\ref{fig_superpos_109}.  The average model dip profile shows a clear
bump of $\sim8$\% amplitude on either side of the dip due to
diffraction, whereas the averages of the actual profiles show no
significant evidence for diffraction sidelobes.  Thus, we conclude
that the fraction $f_{\rm occ}$ of legitimate TNO occultations can be no
larger than $\sim$30\%, otherwise diffraction sidelobes likely would
have been detected.  Again, while this is a clear strike against the
dips being due to TNOs, the limiting statistically significant
constraint that can be set due to the lack of diffraction sidelobes is
not as significant as for the SM1-other events.

(4) {\bf Asymmetry:} A comparison of the simulated with the actual dip
profiles (as in [3]) above, clearly shows a marked asymmetry for the
real dip events.  This is physically implausible if the dips are the
product of occultation events and therefore testifies against a TNO
origin for most of the dip events.  We estimate that the statistical
significance of the asymmetry is $\sim 6\,\sigma$.  Unfortunately,
there is no direct way to use this information to constrain the
fraction of legitimate TNO occultations.  The problem is that we do
not know, a priori, how large the asymmetry is, on average, for
non-TNO dips.  Therefore, we can not tell how `diluted' the non-TNO
events are by potentially real ones.  Nonethless, this marked
asymmetry is another solid indication that few of the dips are the
product of TNO occultations.

(5) {\bf Lack of dips longer than $\sim$1 ms:} From
Fig.~\ref{fig_realwd} we can see that all of the dips, except for a
single event, have RMS widths $\sigma < 1.1$ ms.  In
Section~\ref{sec:data} we described a computer simulation of the
production, detection, and analysis of dips caused by TNO
occultations.  For a relative speed between the {\it RXTE} satellite
and the shadows of the putative TNOs of $v_{\rm rel} = 25$ km s$^{-1}$ we
find that the fraction of recovered simulated dips with $\sigma > 1.1$
ms is $\sim$27\%.  For $v_{\rm rel} = 35$ km s$^{-1}$, $9$\% of the dips
have $\sigma > 1.1$ ms.  We estimate that the average relative
velocity between {\it RXTE} and the shadows of any TNOs was not higher
than $v_{\rm rel} \sim 30$ km s$^{-1}$.  For this speed, $16$\% of the
dips are characterized by $\sigma > 1.1$ ms.  Therefore, if {\em all}
of the dips are the result of TNO occultations the number of
longer-duration dips should be $\sim$30, whereas the observed number
is actually 1.  On the other hand, if only 15\% of the dips are due to
TNO occultations, then we would expect only $\sim 5$ dips with $\sigma
> 1.1$ ms.  This expected number is marginally statistically
consistent, i.e., at $\sim$5\% confidence, with the detection of one
dip with $\sigma > 1.1$ ms.  Therefore, we conclude that the lack of
longer dips allows an upper limit of 15\% to be set on the fraction,
$f_{\rm occ}$, of potentially real TNO occultations.

(6) {\bf Lack of correlation between width and depth:} If the dips
were due to TNO occultations of Sco X-1, we would expect a strong
correlation between the widths of the dips and their depths.  This
results from the fact that diffraction produces shallow occultations
for the smaller size occulters, while it produces deeper more
geometric-shadowing-like occultations for the larger occulters.  As
can be seen from the distribution of dip widths vs. depths in
Fig.~\ref{fig_realwd}, there is no such correlation, with almost all
of the dips confined to a narrow range of widths (between 0.4 and 0.8
ms) and depths that range all the way from 45\% to nearly 100\%.
Thus, the fact that the dips we detect include a significant number,
i.e., $\sim$20\%, that are both narrow ($\sigma < 0.7$ ms) and deep
(minimum normalized count rate below 0.2) whereas only $\sim2$\% of
the `detected' simulated dips (for $v_{\rm rel} \sim 30$ km s$^{-1}$)
are this narrow and deep, indicates that $\lesssim10$\% of the dips
might be due to TNO occultations.  Given the effects of statistical
fluctuations on the observed number of narrow deep dips and the fact
that the simulation is based upon somewhat uncertain parameters, it is
more reasonable to use these numbers to set an upper limit of
$\sim20$\%.

Summarizing the results from approaches (1) through (6) above, we find
limits on the fraction of valid TNOs to be $f_{\rm occ} <11\%$, $< 38\%$,
$< 30\%$, $< Q\%$, $<15\%$, $<20\%$, respectively, where ``Q'' denotes
that a formal limit could not be set, but the approach provides an
important independent indication that the dips are, for the most part,
not the result of TNO occultations.

We believe that the combined upper limit on $f_{\rm occ}$ due to the joint
application of all six approaches is simply the minimum value achieved
by the most sensitive of these, i.e., the constraints cannot be
combined.  The reason, in short, is that the effects we explore serve
only to statistically limit the number of events which could be due to
TNOs rather than to identify specific qualifying events.  Therefore,
our final limit is simply $f_{\rm occ} \lesssim 10\%$.

One might argue, as did \citet{chang07}, that since $\sim10$\% of the
observed dips cannot be formally eliminated as being due to TNOs, they
serve as viable potential candidates for TNO detections.  However, we
argue that if 90\% of the dips can be securely eliminated as TNO
occultations, and there are six different and independent indicators
that point in the direction of a common cause due to cosmic ray
interactions in the detector, then it is most plausible that {\em all}
of the dips have this common origin.

While our results cast serious doubt on whether any true occultation
events have been detected, one cannot yet conclude with a high degree
of confidence that no such events have been detected. Further
investigations of the dip phenomenon and its possible causes would be
of interest.  We are working to obtain a new measurement of, or upper
limit on, the rate of occurrence of occultations of Sco X-1 by
analyzing the data that are being obtained in a new series of {\it
RXTE} observations of Sco X-1 with high-time-resolution information on
VLE events.

\section{Implications for the Population of TNOs}

Given that we detected 203 dips in data that covered 570 ks of
observations with count rates $\gtrsim 55,000$ cts s$^{-1}$, our upper
limit on $f_{\rm occ}$ corresponds to an upper limit of $\lesssim20$
occultations in the 570 ks of observations.  If we adopt a model of
the TNO characteristics, this upper limit of detected TNO occultations
can be used to establish an upper limit on the abundance of small
TNOs.  For this purpose we use the assumptions and results of our
model TNO simulations described above in Section~\ref{sec:data}.

\begin{figure}[t]
\plotone{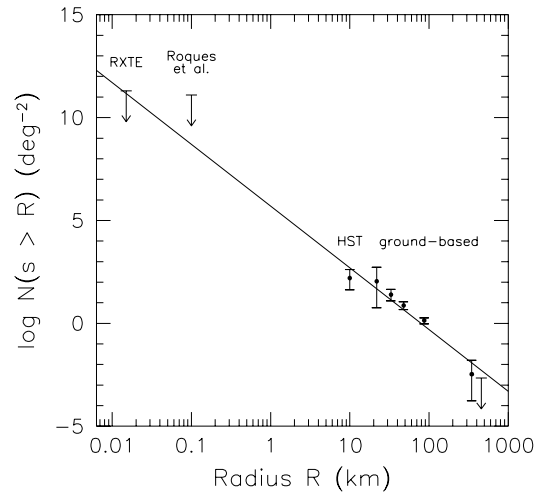}
\caption[Summary of measurements of the populations of small
TNOs]{Summary of measurements of the cumulative size distribution of
small TNOs including the upper limit determined herein from {\it RXTE}
PCA observations.  The {\it HST} and the ground-based optical
measurements to the right of the {\it HST} point have been adapted
from Table 2 of \citet{bern04} by converting limiting R magnitudes to
radii by simply scaling from the {\it HST} limiting magnitude which we
take to correspond to a radius of 10 km at a distance of 40 AU.  We
show error bars that give the ranges of values consistent with Poisson
statistics at confidence levels of 90\%. The measurements reported in
\citet{roq06} are shown as an upper limit. The line represents a
differential size distribution of $\frac{dN}{ds} \propto s^{-4}$.}
\label{fig_tno_pop}
\end{figure}

A TNO of radius $s$ appears to sweep out a solid angle per unit time
\begin{equation}
 \frac{d\Omega}{dt} = \frac{2(s + \delta) v_{\rm rel}}{D^2}
\end{equation}
where $s + \delta$ is the maximum impact parameter for which this body
may produce a detectable dip, $v_{\rm rel}$ is the apparent transverse
velocity of the TNO, and $D$ is the distance from Earth to the TNO.
We find from our simulations of occultations that, on account of
diffraction effects, $\delta \sim 7$ m for the model TNOs in the
relevant size range.  We use $v_{\rm rel} \sim 30$ km/s and $D \sim 40$
AU.  The solid angle swept out in time $\Delta t$ is
\begin{equation}
 \Omega_{sw} = \frac{d\Omega}{dt}\Delta t = \frac{2(s + \delta) v_{\rm rel}}{D^2}\,\Delta t .
\end{equation}
For an ensemble of TNOs of various sizes, the average solid angle
swept out per TNO of radius $s > s_{min}$ is then
\begin{eqnarray}
 \overline{\Omega}_{sw} & = & \frac{1}{N(s > s_{min})}
    \int_{s_{min}}^{\infty}\Omega_{sw}(s)\frac{dN}{ds}\,ds \\
    & = & \frac{2 v_{\rm rel}\Delta t}{D^2}\,(3 s_{min}/2 + \delta)
\end{eqnarray}
where we have assumed that the differential size distribution of TNOs
is given by $\frac{dN}{ds} \propto s^{-4}$ for $s \gtrsim s_{min}$.
In our simulations we find $s_{min} \sim 15$ m.  Using this and the
above values, one obtains
\begin{equation}
 \overline{\Omega}_{sw} \sim 2.8 \times 10^{-14}\, \mathrm{sr} 
                        \sim 9 \times 10^{-11}\, \mathrm{deg}^2 .
\end{equation}
Given that as many as 20 TNOs may have been detected, the upper limit
on the areal density of TNOs is then
\begin{equation}
 N(s \gtrsim 15\,\mathrm{m}) \lesssim \frac{20}{9 \times 10^{-11}\, \mathrm{deg}^2}
       \sim 2 \times 10^{11}\, \mathrm{deg}^{-2} .
\end{equation}

This upper limit may be compared to previous measurements of the size
distribution of TNOs at larger radii.  Figure~\ref{fig_tno_pop}
summarizes the results of surveys of TNOs smaller than 1000 km.
The smallest TNOs which have been securely detected were found in the
Hubble Space Telescope ACS survey reported by \citet{bern04}, in which
3 TNOs of radius $s \gtrsim 10$ km were found in 0.019 square degrees
of sky. We also show the population estimates from ground-based surveys
that are summarized in Table 2 of \citet{bern04}.

\acknowledgments

We are grateful to Jean Swank for helpful discussions on the technical
aspects of the PCA and to Jim Elliot for a number of helpful
discussions.


\begin{thebibliography}{}

\bibitem[Bernstein et al.(2004)]{bern04} Bernstein, G.~M., 
Trilling, D.~E., Allen, R.~L., Brown, M.~E., Holman, M., \& Malhotra, R.\ 
2004, \aj, 128, 1364 

\bibitem[Cash(1979)]{cash79}
Cash, W.\ 1979, \apj, 228, 939 

\bibitem[Chang et al.(2006)]{chang06} Chang, H.-K., King, 
S.-K., Liang, J.-S., Wu, P.-S., Lin, L.~C.-C., \& Chiu, J.-L.\ 2006, \nat, 
442, 660 


\bibitem[Chang et al.(2007)]{chang07} Chang, H.-K., Liang, 
J.-S., Liu, C.-Y., \& King, S.-K.\ 2007, ArXiv Astrophysics e-prints, 
arXiv:astro-ph/0701850 

\bibitem[Jahoda et al.(2006)]{pca06} Jahoda, K., Markwardt, 
C.~B., Radeva, Y., Rots, A.~H., Stark, M.~J., Swank, J.~H., Strohmayer, 
T.~E., \& Zhang, W.\ 2006, \apjs, 163, 401 

\bibitem[Jones et al.(2006)]{tajatel} Jones, T.~A., Levine, 
A.~M., Morgan, E.~H., \& Rappaport, S.\ 2006, The Astronomer's Telegram, 
949, 1 

\bibitem[Jones et al.(2007)]{tajast} Jones, T.~A., Levine, A.~M.,
Morgan, E.~H., \& Rappaport, S.\ 2007, ArXiv Astrophysics e-prints,
arXiv:astro-ph/0612129v2

\bibitem[Roques et al.(2006)]{roq06} Roques, F., et al.\ 
2006, \aj, 132, 819 


\end{thebibliography}
\end{document}